\documentclass[reprint,onecolumn]{revtex4-2}
\usepackage{amsmath,amsfonts,amssymb}
\usepackage{graphicx,wrapfig}
\usepackage[hyphens]{xurl}
\usepackage{hyperref}

\newcommand{\mL}{\mathcal{L}}
\newcommand{\mR}{\mathcal{R}}
\newcommand{\mB}{\mathcal{B}}
\newcommand{\mC}{\mathcal{C}}

\newcommand{\du}{\dot{u}}

\newcommand{\Div}{{\rm div}}

\begin{document}

\title{Spatio-temporal air flow properties in a 3D personalised model of the human lung.}

\author{Jonathan Stéphano}
\affiliation{Université Côte d'Azur, LJAD, Parc Valrose, 06108 Nice, cedex 2}
\author{Michaël Brunengo}
\affiliation{Université Côte d'Azur, LJAD, Parc Valrose, 06108 Nice, cedex 2}
\author{Riccardo Di Dio}
\affiliation{Université Côte d'Azur, LJAD, Parc Valrose, 06108 Nice, cedex 2}
\author{Thomas Laporte}
\affiliation{Université Côte d'Azur, LJAD, Parc Valrose, 06108 Nice, cedex 2}
\author{Benjamin Mauroy}
\affiliation{Université Côte d'Azur, CNRS, LJAD, Parc Valrose, 06108 Nice, cedex 2}
\email[]{benjamin.mauroy@univ-cotedazur.fr}

\begin{abstract}
We propose a multi-scale lung model to investigate spatio-temporal distributions of ventilation variables. Lung envelope and large airway geometries are derived from CT scans; smaller airways are generated using a physiologically consistent algorithm. Tissue mechanics is modeled using nonlinear elasticity under small deformations, coupled with local air pressure from fluid dynamics within the bronchial tree. Airflow accounts for inertia and static airway compliance. Simulations employ finite elements. Using this model, we explore spatio-temporal airflows and shear stresses distributions.
\end{abstract}


\maketitle

\section{Introduction}

The bronchial mucus protects the lung from external aggression by trapping and removing inhaled particulate matter. Mucus is continuously produced and transported toward the esophagus through two primary mechanisms: mucociliary clearance and cough. In pathological conditions where these clearance processes are impaired, mucus stagnation occurs, increasing susceptibility to infections. Chest physiotherapy (CP) is often prescribed to compensate for this dysfunction and facilitate mucus transport. Many common CP techniques involve generating high airflow rates, based on the rationale that such flows interact with mucus and promote its mobilization. However, analyses of this interaction and the resulting mucus dynamics throughout the bronchial tree remain largely empirical.

Mathematical and computational modeling is particularly well suited to studying the lung's internal dynamics, which are challenging to observe in vivo. Recent studies employing Weibel-like models of the bronchial tree have demonstrated that air–mucus interactions can indeed mobilize mucus during chest physiotherapy. Key drivers of these interactions have been identified and characterized, including air wall shear stress, mucus rheology, and airway geometry and structure \cite{mauroy_toward_2015, stephano_wall_2021}. However, these idealized models assume that all pathways from the trachea to the acini are geometrically and biophysically identical, and thus cannot capture spatial inhomogeneities. A significant paradigmatic gap remains in modeling the three-dimensional spatial distributions of the underlying physical processes.

In this abstract, we propose a new multi-scale lung model that addresses this gap under resting ventilation conditions, incorporating key processes such as the coupling between local tissue deformation, airway compliance, and inertial airflow \cite{neelakantan_computational_2022}.
Using this model, we explore the spatio-temporal distributions of air wall shear stresses and airflows within our integrative lung model at rest ventilation.

\section{Methodology}

\subsection{3D geometrical models}
{\bf Geometrical Model.} The model geometry consists in 3D models of the left ($\mL$) and right ($\mR$) lung envelopes, whose walls are denoted $W_p$, and from a 3D model of the upper bronchial tree $\mB$. 
$\mB$ comprises the airway walls $W_a$, the tracheal opening $T$, and $n_o$ terminal openings $(O_i)_{i=1..n_o}$, each located in either the left ($i \in I_L$) or right ($i \in I_R$) lung. 
From each terminal opening $O_i$ of the 3D mesh, we decompose its host lung ($\mL$ or $\mR$) into subdomains $(A_i)_{i=1..n_o}$ via Voronoi tessellation based on the barycenters of the $O_i$.

The thorax is modeled as the convex hull $\mC$ of both lungs dilated by 10$\%$. The pleural cavity is simplified by assuming an elastic medium fills the space between the lungs and thoracic wall.

Geometries are represented using surface and volumetric meshes.
$\mL$, $\mR$ and $\mB$ are reconstructed from patient CT scans \cite{laporte_du_2023}. 
Here, segmentations from the \href{https://luna16.grand-challenge.org/}{\it LUNA16 challenge} were used, with surface meshes generated from 3D masks via \href{https://www.cgal.org/}{\it CGal} and volumetric meshes obtained using \href{https://www.meshlab.net/}{\it MeshLab}. 
Mesh resolution is controlled throughout the process.\\

{\bf Fluid Mechanics.} Airflow in $\mB$ is governed by the nonlinear incompressible Navier–Stokes equations, solved using 3D finite elements and the method of characteristics. Atmospheric pressure is the reference. No-slip conditions are imposed on $W_a$. The extrathoracic airways are modeled via a resistance $R_T = 50 \, 000$ Pa·m$^{-3}$·s, imposing $-p_{T,\text{mean}} = R_T f_T$, where $p_{T,\text{mean}}$ is the mean pressure at $T$ and $f_T$ the tracheal flow. The associated Lagrange multiplier corresponds to a uniform pressure applied over $T$. Similarly, for each terminal opening $O_i$, flow is constrained to equal the volume change rate of its associated lung region $A_i$, yielding uniform pressures $p_{O_i}$ as Lagrange multipliers.\\

{\bf Tissue Mechanics.} Lung tissue mechanics is modeled using small-strain elasticity with Poisson's ratio $0.3$ and a nonlinear local deformation-dependent Young's modulus, calibrated from literature's static compliance data.
Lamé parameters $\mu$ and $\lambda$ are deduced from these quantities.
Lung density accounts for its 90 $\%$ air, 10 $\%$ tissue composition, $\rho = 100$ kg.m$^{-3}$. 
The stress–strain relation incorporates both tissue deformation and local air pressure \cite{brunengo_optimal_2021}:
\begin{equation}
\sigma(u,\dot{u}) = 2\mu(\nabla u) \ \epsilon(u) + \lambda(\nabla u) \ \mathrm{tr}(\epsilon(u)) I - p(\dot{u}) I
\label{eq1}
\end{equation}
where $u$ is displacement and $p$ is local air pressure. Air pressure at a point $x$ within region $A_i$ combines the pressure at its feeding terminal opening $O_i$ and the pressure drop through the distal airway subtree. 
Modeling these pressures requires generating physiologically consistent subtrees peripheral to $\mB$, see section \ref{smallAirways}.\\

\begin{figure}[t!]
\centering 
a \includegraphics[height=4cm]{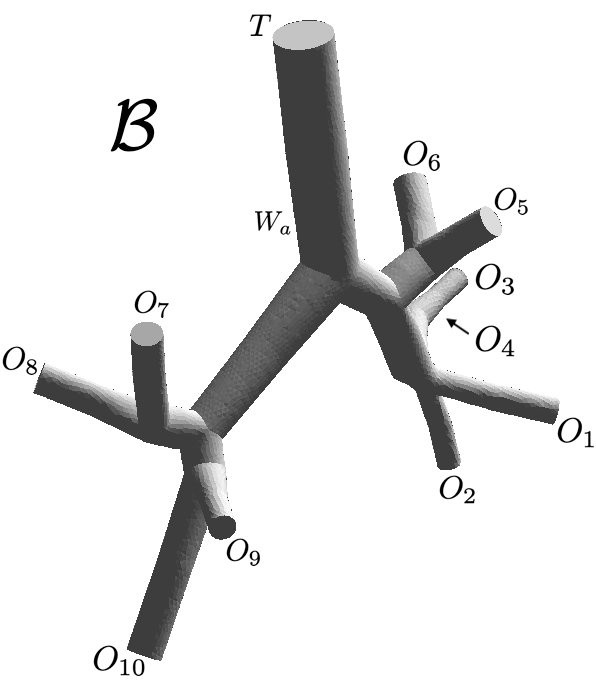}
b \includegraphics[height=4cm]{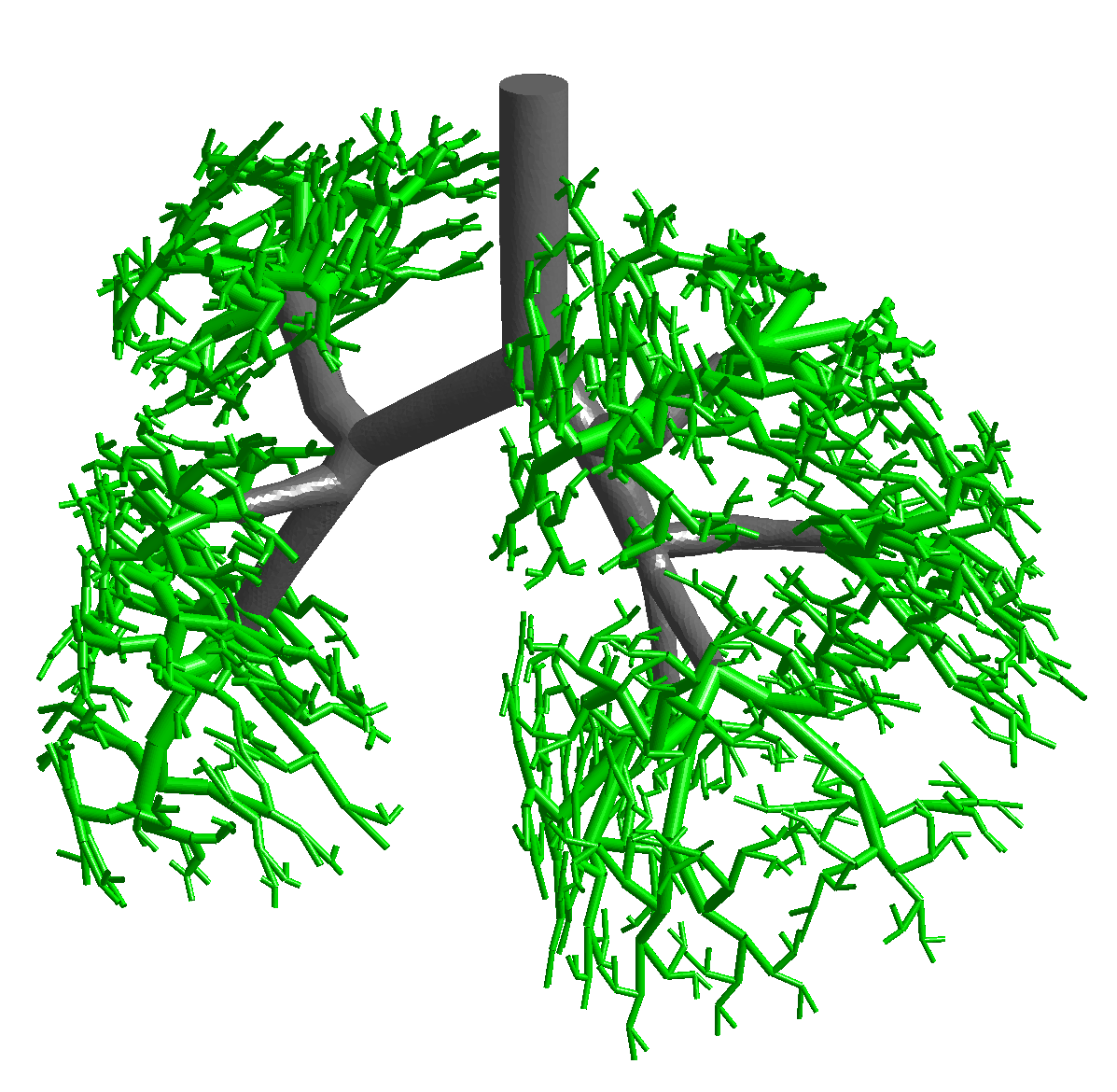}
c \includegraphics[height=4cm]{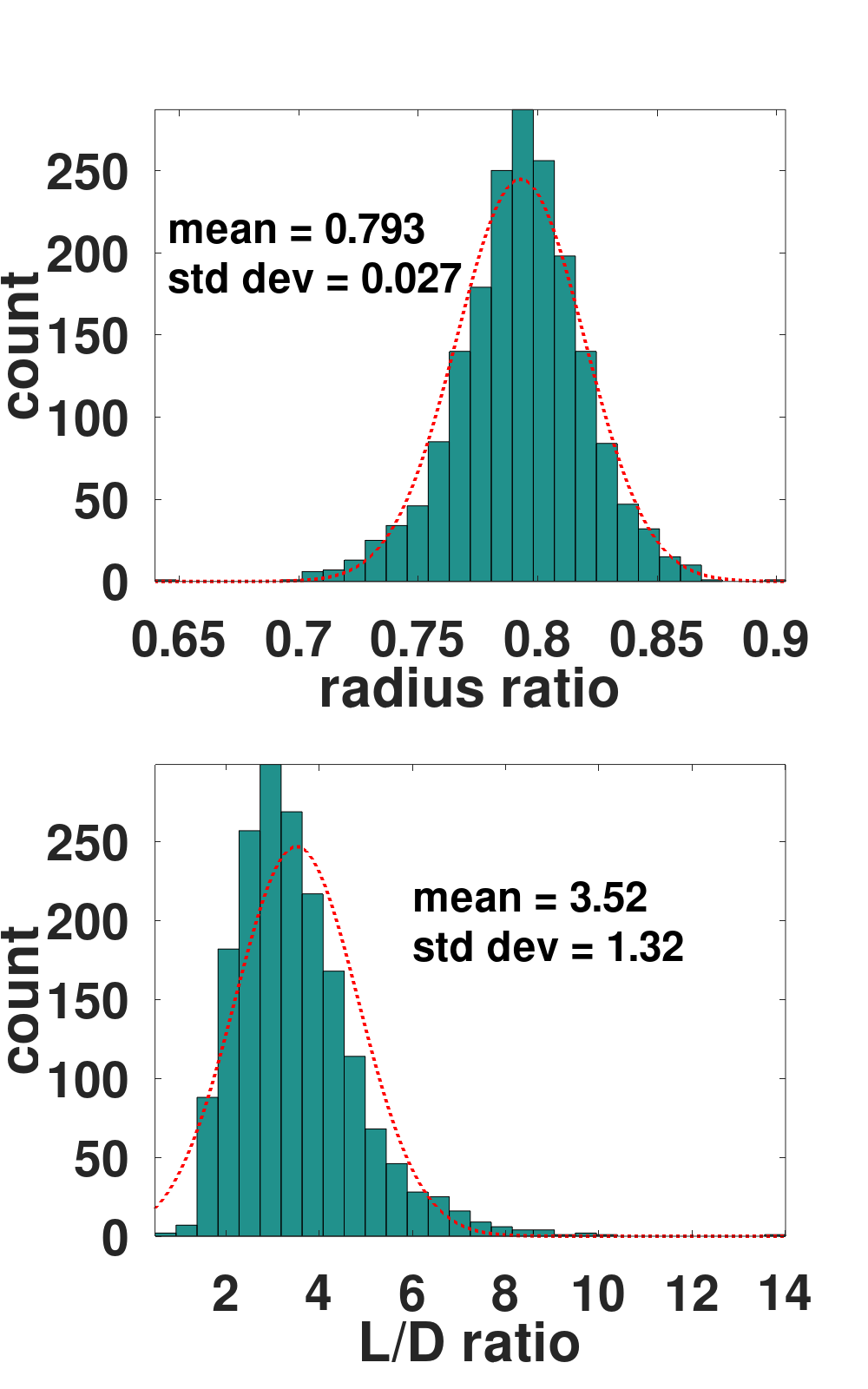}
d \includegraphics[height=4cm]{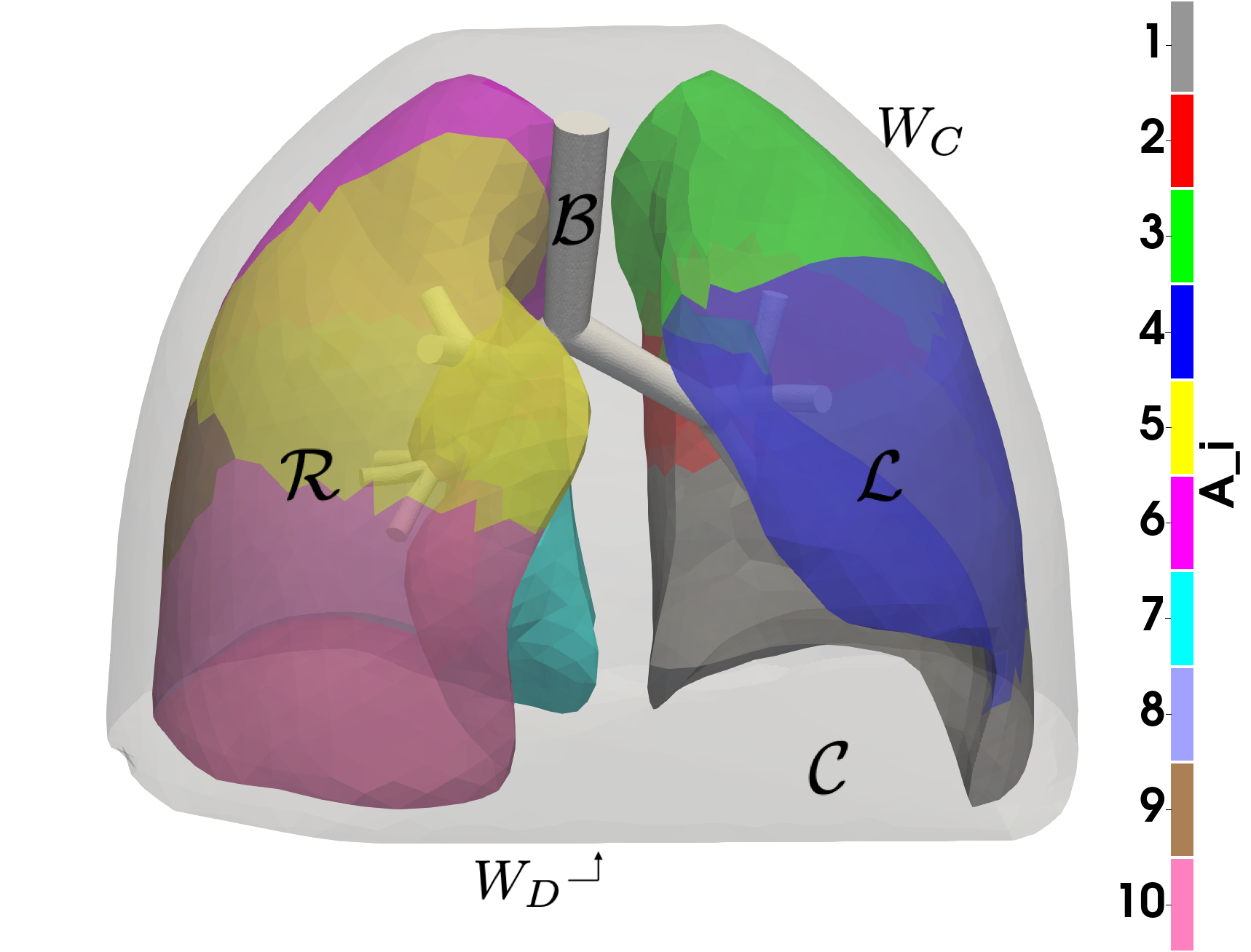}
\caption{Example of lung model.  {\bf a}: Large airways reconstructed from ct-scans.
{\bf b}: Airways in gray are reconstructed from CT scans and include 10 terminal openings $(O_i)_{i=1..10}$. Each opening connects to a green airway subtree generated by our algorithm.
{\bf c}: Size statistics for airways generated by the reconstruction algorithm (green in panel a). Top: airways daughter-to-mother diameter ratio distribution. Bottom: airways length-to-diameter ratio distribution. Both distributions are consistent with observed data.
{\bf d}: Reconstructed 3D model: large airways (gray scale: air velocity), left and right lungs (colors: index of $(A_i)_{i=1..10}$ decomposition), and their idealized thoracic envelope.}
\label{fig1}
\end{figure}

{\bf Boundary conditions.} At the intersections of $\mB$ with the walls of $\mL$ and $\mR$, and on the spine region at the back of $\mC$, we assume no displacements ($u=0$).
A time-dependant pressure is applied to the lower part $W_D$ of $\mC$ to mimic the diaphragm-induced stress.
Half that pressure is applied on the rest of $\mC$ wall to mimic coastal muscles-induced stress.
The pressure amplitude is calibrated to obtain a physiologic rest lung tidal volume.
The pressure time dependance is a smoothed Heaviside, with a $1.3$ s active inspiration and a $2.6$ s passive expiration, see \cite{brunengo_optimal_2021}.

\subsection{Small airways models}
\label{smallAirways}

\begin{figure}[t!]
\centering 
a \includegraphics[height=3.5cm]{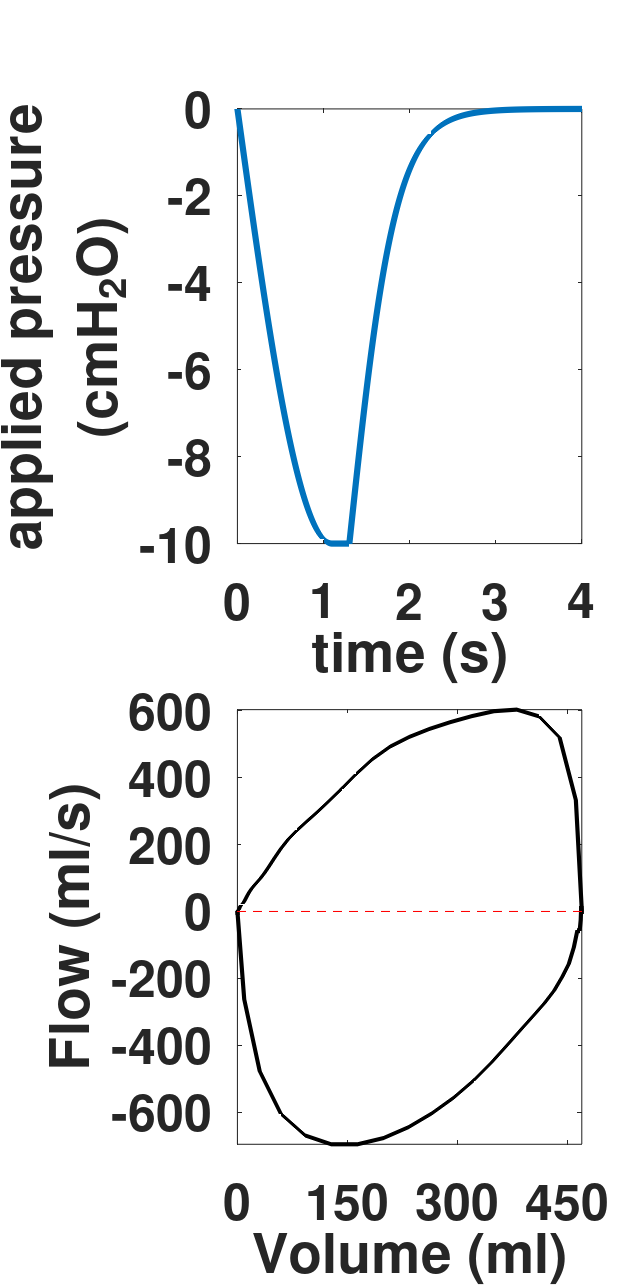}
b \includegraphics[height=3.5cm]{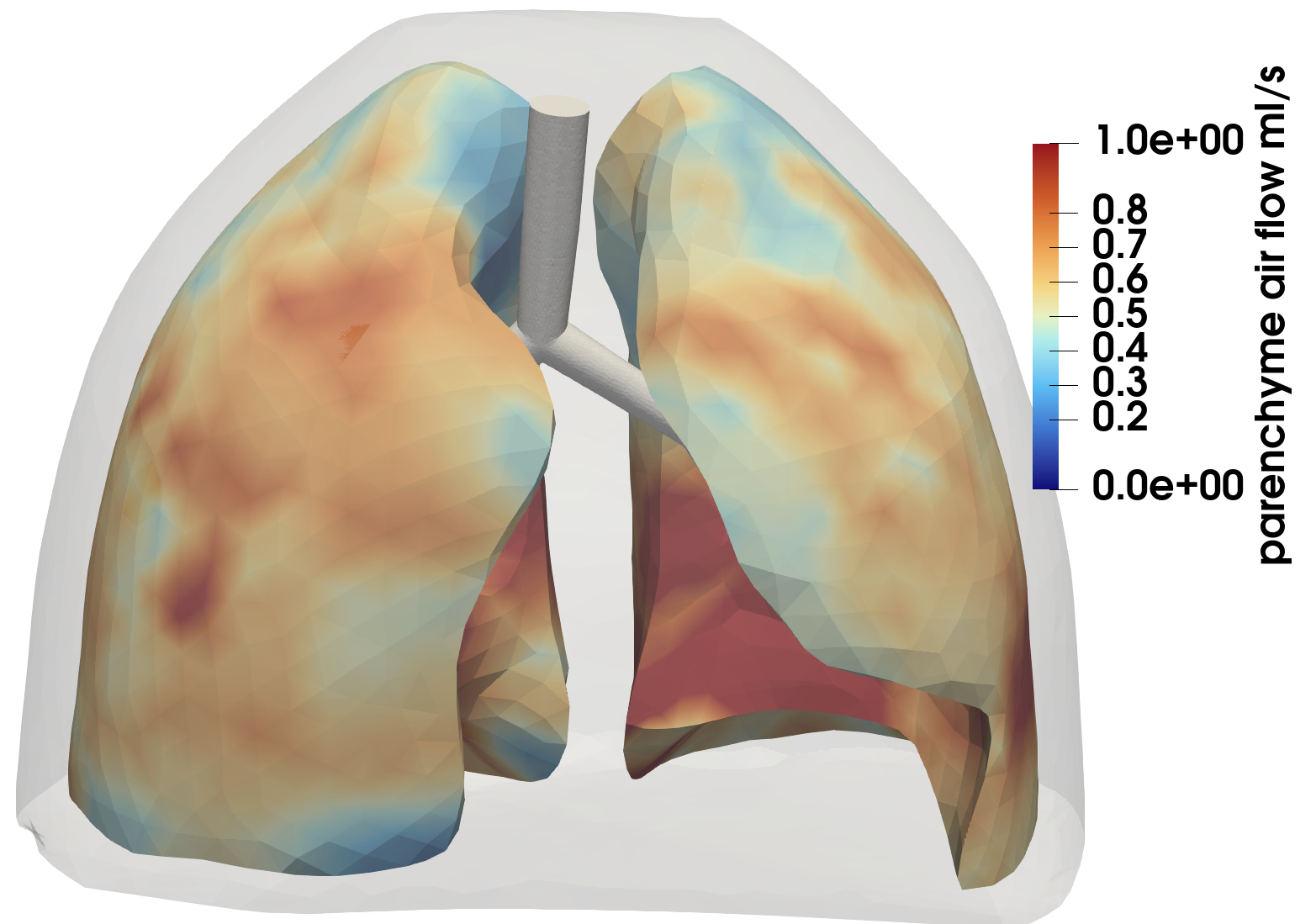}
c \includegraphics[height=3.5cm]{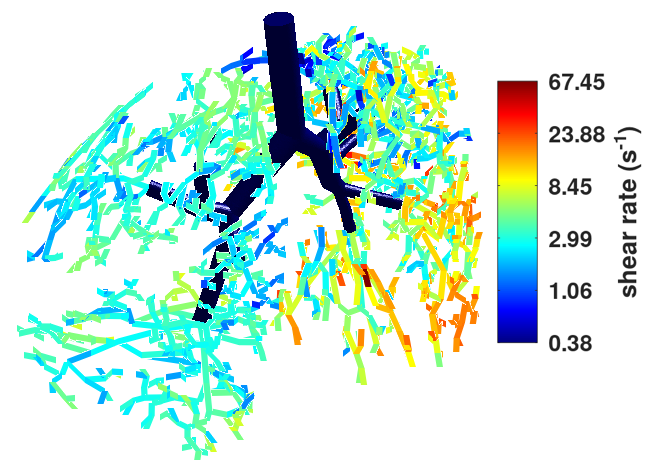}
d \includegraphics[height=3.5cm]{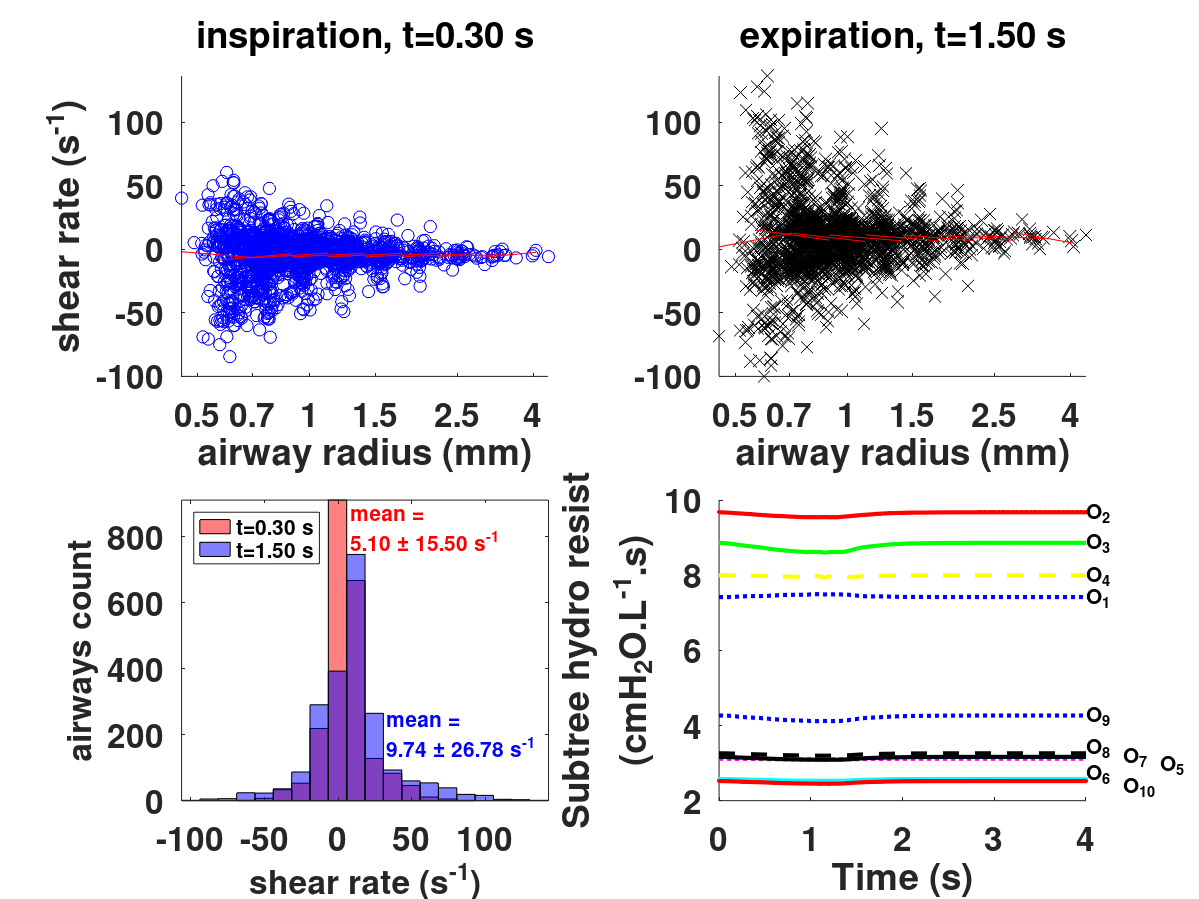}
\caption{Lung model simulation at rest.
{\bf a}: {\it Top}: Applied thoracic pressure over time. {\it Bottom} : Corresponding airflow–volume loop. Hydrodynamic resistance: $1.09 \times 10^5$ Pa·m$^{-3}$·s ($1.09$ cmH$_2$O·L$^{-1}$·s); compliance: $0.475 \times 10^{-6}$ m$^3$·Pa$^{-1}$ ($0.0475$ L·cmH$_2$O$^{-1}$).
{\bf b}: Front view of the lung model, color-coded for local airflow. The complex pattern arises from the heterogeneous hydrodynamic resistance of the paths feeding each peripheral region $B_{i,k}$. 
{\bf c}: Mean shear rates over one cycle in the reconstructed airways (rear view).
{\bf d}: Shear rate distributions at peak flow. {\it Top}: Shear rate distribution vs. airway radius. {\it Lower right}: Overall shear rate distribution at peak inspiration (red) and expiration (blue). {\it Lower left:} Time-varying hydrodynamic resistance of the ten subtrees.}
\label{fig2}
\end{figure}

{\bf Small Airways Modeling.} CT scans resolve only larger airways, so smaller ones are generated algorithmically. We use a 0D representation where airways are defined by radius and endpoint positions, assuming a purely bifurcating distal tree.

Each lungs subregions $A_i$ is recursively subdivided. At each step, a region is split by the plane containing the end $e$ of its associated branch (radius $r_m$) and the orientation vectors of its parent and sister branches. Two new airways are created, sprouting from $e$ and oriented toward the barycenters of the two subregions. Airway length is a fraction $\alpha = 0.6$ of the distance from $e$ to these barycenters. Radii $r_1$ and $r_2$ follow Hess-Murray law: $r_m^3 = r_1^3 + r_2^3$ with $(r_1/r_2)^3$ being the volumetric ratio of the corresponding two new subregions. Subdivision continues until subregion volume falls below the typical volume of an acinus or twice the local mesh element size. The result is a space-filling tree where each $A_i$ is partitioned into terminal subregions $(B_{i,k})_{k=1..m_i}$.\\

{\bf Deep Airways.} When the above algorithm stops before reaching terminal bronchioles, a symmetric distal tree is appended. Radii and lengths decrease by factor $h = 2^{-1/3}$ per generation until radius reaches $r_{\mathrm{tb}} \approx 0.25$ mm. Within acini (beyond terminal bronchioles), $h = 1$ to preserve physiological architecture. Airways of identical size are assumed geometrically and physically equivalent.\\

{\bf Airway Mechanics.} Small and deep airways compliance is modeled via static transmural pressure–area laws \cite{lambert_computational_1982}, interpolated to arbitrary airway sizes.\\

{\bf Fluid Mechanics and coupling with tissue mechanics.} Air fluid dynamics in small and deep airways is treated as 0D and includes some inertia. Pressure drop $\Delta p_a$ across airway $a$ of radius $r_a$ and length $l_a$ is \cite{stephano_wall_2021}:
$$
\begin{array}{l}
\Delta p_a = R_a\left(f_a\right) f_a\\
\text{ with } R_a(f_a) = 
\left\{ 
\begin{array}{ll}
\frac{16 l_a \mu}{\pi r_a^4}&\text{if ${\rm Re_a} < 300$}\\
\frac{8 l_a \mu}{\pi r_a^4} + \frac{16 l_a \rho f_a}{600 \pi^2 r_a^5}&\text{if ${\rm Re_a} \geq 300$}
\end{array}
\right.\\
\text{ and } {\rm Re_a} = \frac{4 \rho f_a}{\mu \pi r_a}
\end{array}
$$
with $\mu$ air viscosity, $\rho$ air density. Radius $r_a$ depends on transmural pressure, coupling airway mechanics to air flow in the airway and surrounding tissue pressure.

For a subtree rooted at $O_i$, pressure drop to any terminal subregion $B_{i,k}$ is:
$
\Delta p_i (x) = \left( \mR_i(F_i) F_i \right)_k \text{ if $x \in B_{i,k}$ }
$, 
where $F_i = (f_k^{(i)})$ are flows through terminal airways. Flows arise from tissue deformation:
$$
f_k^{(i)} = \frac{d}{dt} \left({\rm vol}\left( B_{i,k} \right)\right) \simeq -\int_{B_{i,k}} \Div (\du) dx
$$ 
under small strains. $\mathcal{R}_i$ is the generalized resistance matrix of the subtree $i$, constructed from the path matrix $P_i$ (matrix with entries $p_{a,b}=1$ if airway $b$ lies on path from trachea to terminal airway $a$, $p_{a,b} = 0$ otherwise) and airway resistances diagonal matrix $R_{v,i}$: $\mathcal{R}_i = P_i R_{v,i} P_i^T$.
This matrix generalises to any bifurcating tree the resistance matrix originally defined for a symmetric bifurcating tree \cite{grandmont_viscoelastic_2006}. 

Local air pressure in equation (\ref{eq1}) is then:
\begin{equation}
p(x) = p_i - \left( \mathcal{R}_i(F_i) F_i \right)_k \quad \text{for } x \in B_{i,k}
\label{eq2}
\end{equation}
with $p_i$ the pressure at $O_i$ from the 3D Navier-Stokes flow solution.\\

{\bf Numerical simulations.}
We developed an automated pipeline that takes as input the meshes of the bronchial tree and lung envelope, and outputs lung displacements along with air velocities and pressures in the airways. A postprocessing tool extracts derived quantities such as airway wall shear stresses and regional flow distributions.

The code integrates \href{https://freefem.org/}{\it FreeFem++} (finite elements), \href{https://www.meshlab.net/}{\it MeshLab} (surface meshing), and \href{https://gmsh.info/}{\it GMSH} (volumetric meshing). 
The numerical scheme is semi-implicit: airway deformation and the strain-dependent Young's modulus are treated explicitly. 
The time step adapts to maintain CFL numbers below one for stability. 
The implicit part involves inverting a large block-structured matrix, which is decomposed into smaller subsystems. 
Simulations presented in this abstract run in approximately one day on four processors of a Mac Studio M2 with 192 GB memory.

\section{Results and conclusions}

Our model provides access to internal biophysical variables of the lung and their spatio-temporal dynamics. It was tested at rest ventilation (tidal volume: 469 ml) using the geometry shown in Fig. \ref{fig1}. 
A total of 2154 airways were reconstructed (Fig. \ref{fig1}b, green), with diameters ranging from 0.90 to 8.79 mm. The applied thoracic ventilation signal and the resulting flow-volume curve are presented in Fig. \ref{fig2}a (flow is negative during inspiration).

As expected, predicted airflow and shear rate patterns exhibit significant complexity (Fig. \ref{fig2}b–c). Airflow distribution depends on the path from the trachea to each peripheral region. Due to the intricate geometry, spatially close regions experience different pressures and flows. Mean absolute shear rates are higher during expiration than inspiration (Fig. \ref{fig2}d), which can be explained by: 1/ the time-asymmetric ventilation signal and flow-volume relationship (Fig. \ref{fig2}a); and 2/ airway expansion during inspiration, which reduces subtrees resistances, while expiration leaves airways nearly unchanged (Fig. \ref{fig2}d, lower right). Shear rates span a wide range of positive and negative values, with increasing variability in smaller airways (Fig. \ref{fig2}d). Retrograde airflows relative to the global flow direction suggest local air redistributions between lung compartments.

This analysis is limited to resting conditions due to the small deformation hypothesis. Nevertheless, it shows the potential of multiscale lung models to reveal complex spatiotemporal patterns, even at rest. Applying highly heterogeneous thoracic pressures, such as in chest physiotherapy (CP), may enhance airflow and shear rate heterogeneity, potentially aiding mucus mobilization. However, the complexity of flow patterns and possible side effects (retrograde flows) could interfere with therapeutic goals. Future work will aim to clarify the relationship between tree structure and airflow patterns, incorporate large deformations, and improve the model validation using Electrical Impedance Tomography data.

\end{document}